# Observation of Ultrafast Coherent and Incoherent Spin Torques via Terahertz Spin-Hall Magnetoresistance

Peter Kubaščík[1], Richard Schlitz[2], Oliver Gueckstock[3], Oliver Franke[3], Miina Leiviskä[4], Martin Borchert[5], Gerhard Jakob[6], Kamil Olejník[4], Andrej Farkaš[1,4], Zdenek Kašpar[1,4], Jiří Jechumtál[1], Martin Bušina[1], Eva Schmoranzerová[1], Petr Němec[1], Yizheng Z. Wu[7,8], Georg Woltersdorf[9], Mathias Kläui[6], H. Reichlová[4], Piet W. Brouwer[3], Sebastian T. B. Goennenwein[2], Tobias Kampfrath[3], Lukáš Nádvorník[1,*]

1. Faculty of Mathematics and Physics, Charles University, 121 16 Prague, Czech Republic
2. Department of Physics, University of Konstanz, Konstanz, Germany
3. Department of Physics, Freie Universität Berlin, 14195 Berlin, Germany
4. Institute of Physics, Czech Academy of Sciences, Prague, Czech Republic
5. Max Born Institute for Nonlinear Optics and Short Pulse Spectroscopy, 12489 Berlin, Germany
6. Institut für Physik, Johannes Gutenberg-Universität Mainz, 55128 Mainz, Germany
7. Department of Physics, State Key Laboratory of Surface Physics, Fudan University, Shanghai 200433, China
8. Shanghai Research Center for Quantum Sciences, Shanghai 201315, China
9. Institut für Physik, Martin-Luther-Universität, Halle, Germany

[*] E-mail: lukas.nadvornik@matfyz.cuni.cz

## Abstract

Spin Hall magnetoresistance (SMR) is a cornerstone effect in spintronics, enabling the electrical readout of magnetic states through spin-orbit coupling at interfaces. Despite its widespread use, experimental studies of SMR are limited to static (DC) or low-frequency (GHz) regimes, leaving the ultrafast dynamics of interfacial spin transport in the terahertz (THz) frequency window entirely unexplored. Here, we report the observation of SMR spanning from DC to 1.5 THz in the model SMR system YIG/Pt, bridging the divide between electronic transport and THz optics. We demonstrate that the SMR response exhibits a pronounced low-pass behavior, decreasing significantly with frequency and vanishing above 1.5 THz. Through a dynamic model, we reveal that this unexpected spectral decay arises from a fundamental competition between two distinct interfacial spin-magnetization coupling mechanisms: the frequency-independent coherent spin torque, and an incoherent torque mediated by thermal magnons. Our analysis shows that the reduced THz SMR signal reflects the nonzero response time of the magnon chemical potential, which facilitates the increased outflow of incoherent magnons through the interface at high frequencies. These results resolve the microscopic nature of spin-torque generation at ultrafast timescales. Crucially, this work establishes THz SMR as a powerful, non-contact methodology capable of probing spin-magnon coupling in magnetic thin films, providing the essential tool for advancing the next generation of spintronic devices, particularly those based on high-frequency antiferromagnets and emerging altermagnetic materials.

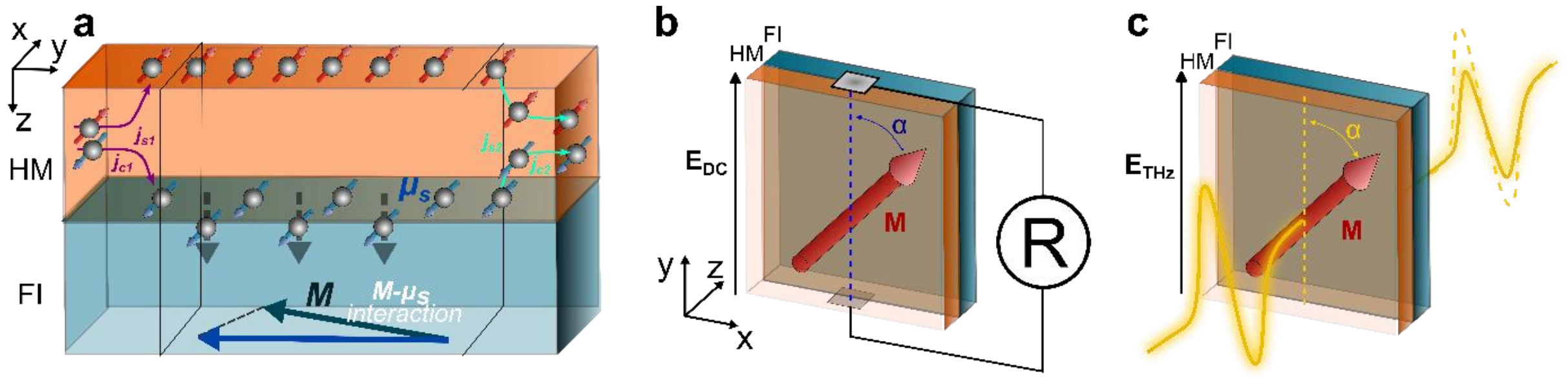

**Fig. 1 | Concept of DC and THz SMR. a** An illustration of the microscopic mechanism of SMR. First, the electric charge current $\boldsymbol{j}_{c1}$ in HM by an applied electric field $\boldsymbol{E}$ is converted into a transverse spin current $\boldsymbol{j}_{s1}$ via the SHE, which leads to a spin accumulation $\boldsymbol{\mu}_s$ at both HM interfaces (circles). Second, $\boldsymbol{\mu}_s$ is partly reduced due to spin relaxation and spin outflow into FI via various interactions with $\boldsymbol{M}$, which generally depend on the $\boldsymbol{M}$ orientation. Finally, the remaining $\boldsymbol{\mu}_s$ leads to a secondary diffusive spin current $\boldsymbol{j}_{s2}$ and, thus, charge current $\boldsymbol{j}_{c2}$ via the inverse SHE, which causes a $\boldsymbol{M}$-dependent variation of the overall sample resistance $\boldsymbol{R}$. **b** Schematic of electrical contact-based transport setup for DC SMR measurement of a FI|HM stack of a ferrimagnetic-insulator layer FI and a heavy-metal layer HM. The in-plane magnetization $\boldsymbol{M}$ of FI and the applied electric field $\boldsymbol{E}_{DC}$ define the angle $\boldsymbol{\alpha}$. **c** Free-space THz setup analogous to **a** with the electric field $\boldsymbol{E}_{THz}$ of a THz pulse, which is detected after transmission through the sample as the electro-optic signal $\boldsymbol{S}(\boldsymbol{\alpha})$.

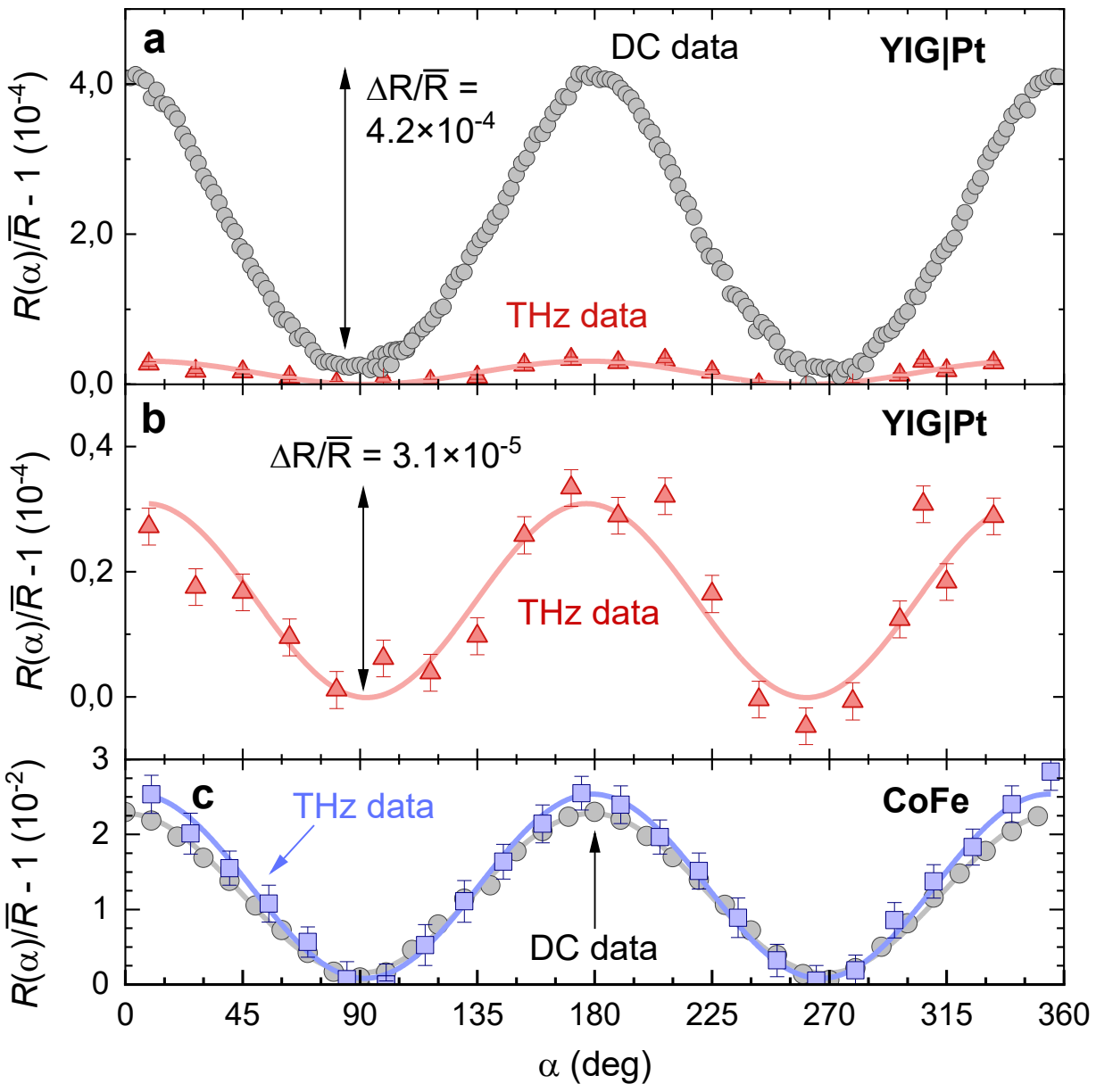

**Fig. 2 | Amplitude and $\alpha$-dependence of the magnetoresistance signals. a** SMR signal $R(\alpha)/\overline{R}-1$ from YIG|Pt(5 nm) in the DC (grey circles, Fig. 1b) and THz experiments (red triangles, Fig. 1c). **b** Magnified THz signals of **a**. In all panels, the curves are vertically offset to make the absolute minima zero. **c** AMR reference signal $R(\alpha)/\overline{R}-1$ of the control sample CoFe as a function of $\alpha$ from the DC [grey circles] and THz experiments [blue squares].The solid lines are fits $\propto \cos 2\alpha$ [see Eq. (1)], and arrows indicate the full contrast $\Delta R/\overline{R}$.

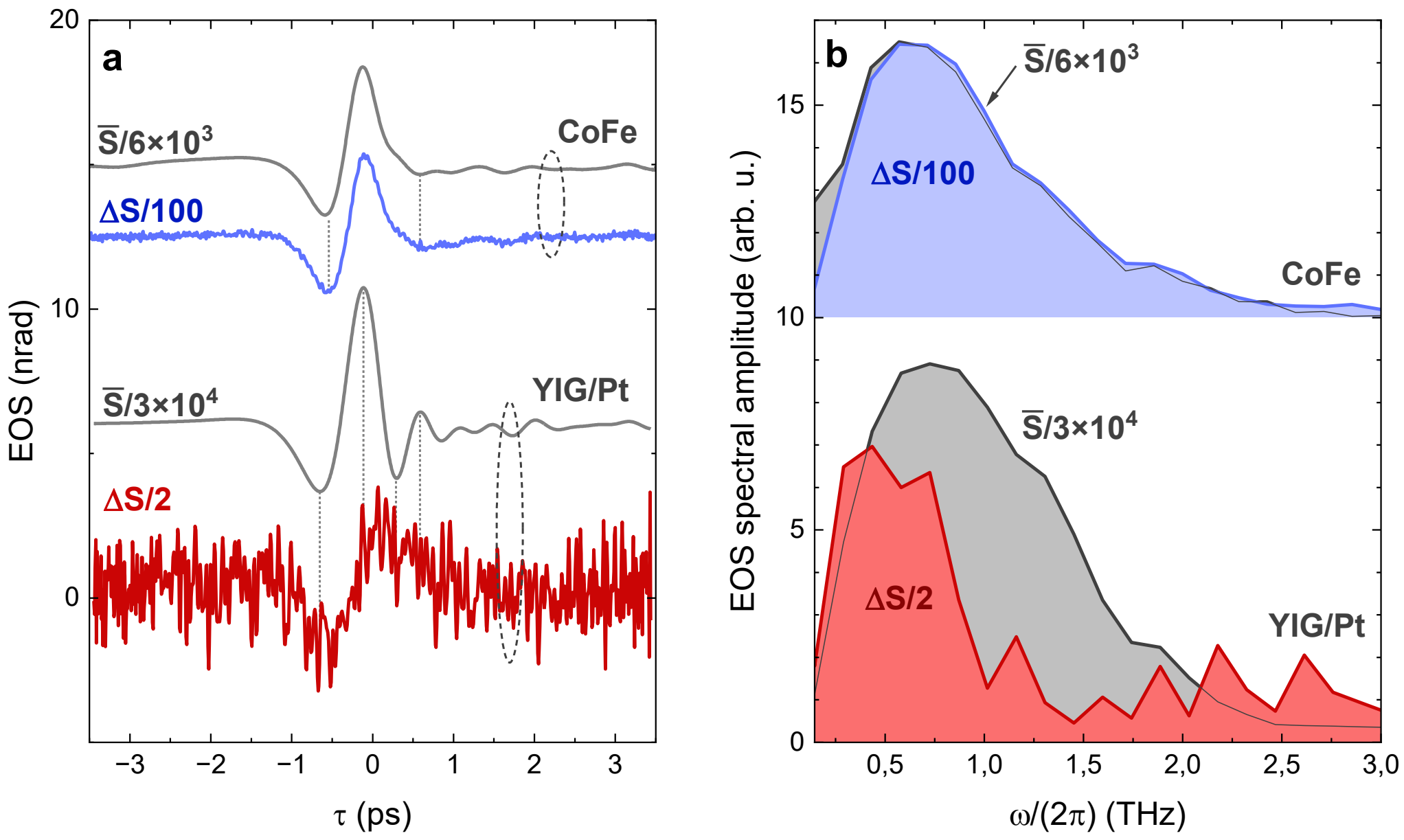

**Fig. 3 | THz differential signals. a** THz waveforms of the magnetization-independent average signals $\overline{\boldsymbol{S}} = (\boldsymbol{S}_{\parallel} + \boldsymbol{S}_{\perp})/2$ (grey curves, rescaled by indicated factors) and the magnetization-dependent differential signals $\boldsymbol{\Delta S} = \boldsymbol{S}_{\parallel} - \boldsymbol{S}_{\perp}$ for YIG|Pt(5nm) (red) and the CoFe reference (blue). Corresponding pairs of $\overline{\boldsymbol{S}}$ and $\boldsymbol{\Delta S}$ for YIG|Pt and CoFe are labeled by dashed ellipse. Dotted lines are a guide for the eye. **b** Amplitude spectra of the signals in **a**, rescaled by the indicated factors. Curves are shifted vertically for clarity.

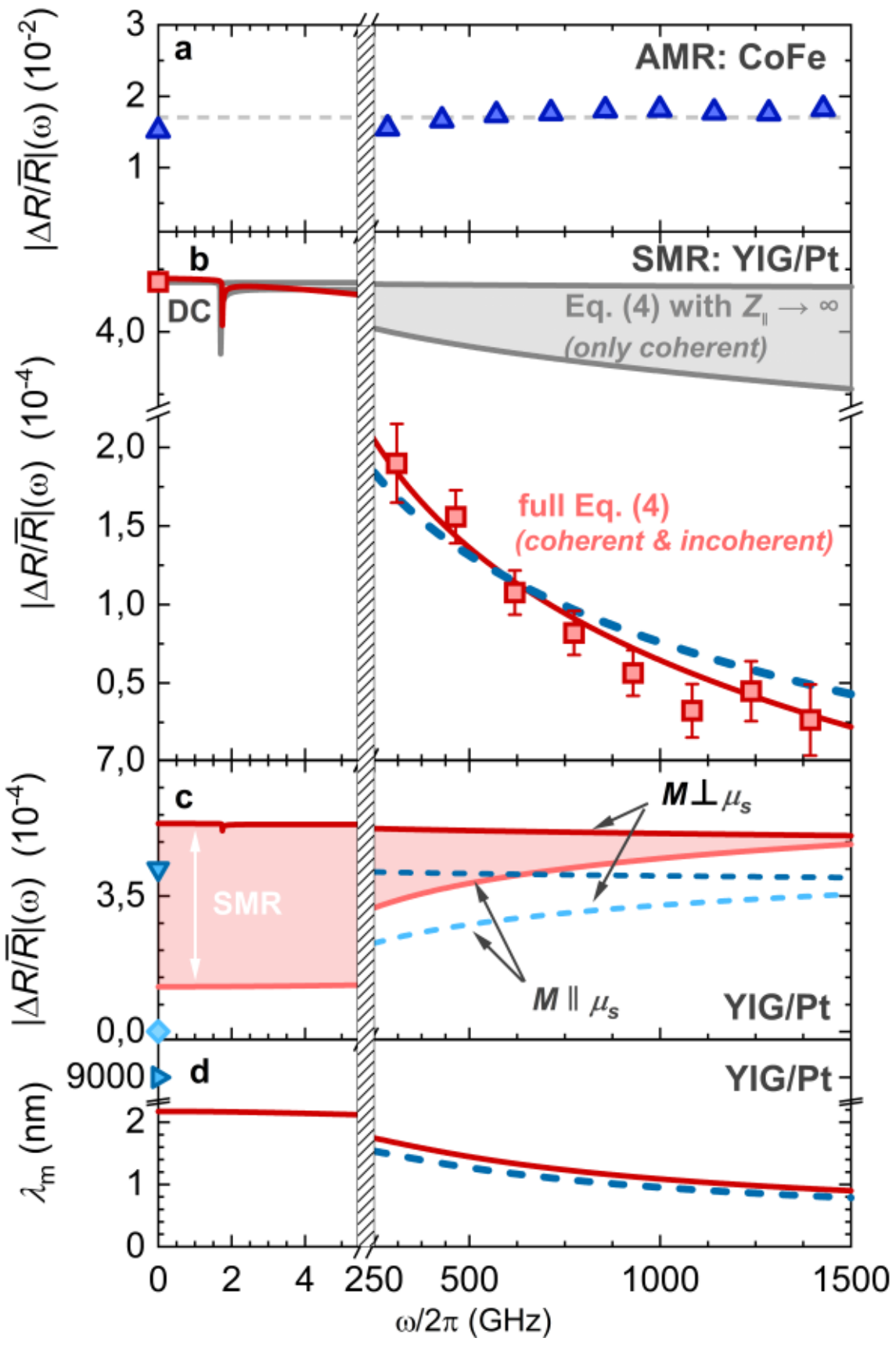

**Fig. 4 | MR contrast vs frequency 0-1.5 THz. a** AMR contrast $|\Delta R/\overline{R}|$ of the reference sample CoFe (blue triangles) as a function of frequency $\omega/2\pi = 0\text{-}1.5$ THz. The DC value is the electrical measurement at $\omega = 0$. **b** Same as **a**, but for the SMR contrast $|\Delta R/\overline{R}|$ of YIG|Pt(5 nm) (red rectangles). Note the split vertical axis. Solid lines are model results [Eq. (4)] for the case of exclusively coherent torque ($Z_\perp$, transverse; grey curves and shaded area) and for the case of coherent ($Z_\perp$, transverse) plus incoherent ($Z_\parallel$, longitudinal) torque contributions relative to the magnetization direction of FI (red curve). Blue dashed curve is the modified model with fixed $\lambda_m(\omega = 0) = 9\ \mu\text{m}$ from Ref. [25]. **c** Coherent and incoherent contributions, $\xi\,\text{Re}[1/Z_\perp(\omega)]$ (the case of $M \perp \mu_s$) and $\xi\,\text{Re}[1/Z_\parallel(\omega)]$ (the case of $M \parallel \mu_s$), for the full model (solid curves) and the modified model (dashed curves). The SMR contrast is the difference of the two contributions. **d** The diffusion length $\lambda_m(\omega)$ of the magnon accumulation in FI for the full and modified models (solid and dashed curves). Note the fixed value of $\lambda_m$ at $\omega = 0$.

The transport of angular momentum by magnons represents a promising pathway to information processing free from Joule heating associated with charge flow [1] and various techniques, many of which require large scale facilities, are employed to study the magnonic spin currents [2]. To study spin transport in insulating ferro- and antiferromagnetic materials via electrical means, the spin Hall magnetoresistance (SMR) [3,4,5] has established itself as a powerful and versatile tool. An SMR experiment consists of a magnetic layer which is interfaced with a non-magnetic metal where a spin current is generated via the spin Hall effect. The resulting change in resistance, as a function of the magnetic state of the adjacent magnetic layer, is then measured. SMR [6] exhibit a characteristic dependence of the resistance

$$R(\alpha) = \overline{R} + \frac{\Delta R}{2}\cos(2\alpha) \tag{1}$$

on the in-plane angle $\alpha$ between the magnetic order parameter (magnetization $\boldsymbol{M}$ or Néel vector) and the electrical current driven by the applied electric field $\boldsymbol{E}_{\mathrm{DC}}$. Here, $\overline{R} = [R(0°) + R(90°)]/2$ is the average of the resistances for $\alpha = 0°$ and $90°$, and $\Delta R = R(0°) - R(90°)$ is their difference.

SMR is widely studied in bilayer heterostructures comprising a heavy-metal (HM), such as platinum or tungsten, and a ferrimagnetic-insulator layer (FI). The prototypical FI material is yttrium iron garnet $Y_3Fe_5O_{12}$ (YIG), which has been the subject of seminal works on SMR [5,6,7,8,9,10,11], alongside materials like GdIG [12]. Crucially, SMR was studied also in antiferromagnets, such as the insulator NiO [13] and the emerging altermagnetic candidates such as α-$Fe_2O_3$ [14] and $Ba_2CoGe_2O_7$ [15]. Furthermore, SMR has also been studied in metallic ferromagnets, such as CoFeB [16,17,18], in which case the both SMR and anisotropic magnetoresistance (AMR) [5,6,7] can contribute to the measured voltage and it needs to be separated carefully [19,20].

The established microscopic framework of SMR in FI|HM stacks [3,21,22,23,24] relies on three concerted spintronic processes (see Fig. 1a): First, the electron current in HM creates a spin accumulation $\boldsymbol{\mu}_s$ at the HM|FI interface via the spin Hall effect (SHE). Second, $\boldsymbol{\mu}_s$ is reduced by a spin outflow into FI through various $\boldsymbol{\mu}_s$-$\boldsymbol{M}$ coupling channels mediated by spin torques [3,17], the efficiency of which depends on the relative orientation of $\boldsymbol{M}$ and $\boldsymbol{\mu}_s$. Third, the backflow of spin due to the $\boldsymbol{M}$-dependent gradient of $\boldsymbol{\mu}_s$ along the stack normal ($z$ axis in Fig.1a) is converted into an in-plane charge current via the inverse SHE, which leads to an $\boldsymbol{M}$-dependent variation of the electrical resistance $R$.

Most experimental studies have focused on the transverse torque-type $\boldsymbol{\mu}_s$-$\boldsymbol{M}$ coupling, where $\boldsymbol{\mu}_s$ is perpendicular to $\boldsymbol{M}$ ($\alpha = 0°$) (Fig. 1a). However, recent advances suggest that additional coupling channels, specifically those mediated by longitudinal incoherent magnons [25,26,27,28,29,30], may play a important role, particularly in insulating systems. It is, however, challenging to experimentally isolate the contributions of these channels. Theory work suggests [23,24] that these intertwined channels should exhibit markedly distinct dynamics in the terahertz (THz) range. Yet, to date, SMR has only been measured at frequencies below $4$ GHz [11]. Consequently, the efficiency of ultrafast magnon excitation by transient spin accumulation and the frequency-dependent nature of spin-to-magnon coupling remain largely unexplored.

In this work, we experimentally bridge this frequency gap by measuring THz SMR in the range of 0–1.5 THz, utilizing a combination of DC transport and THz time-domain spectroscopy on the model YIG|Pt bilayer system. The observed spectral dependence of the SMR exhibits a pronounced low-pass filter characteristic. Comparison with our dynamic SMR model provides the experimental evidence that the interfacial spin flow is governed by a competition between two distinct processes: a nearly frequency-independent contribution from coherent magnons, and a strongly frequency-dependent contribution from incoherent magnons. We demonstrate that the transport of the latter becomes efficient at frequencies exceeding the magnon population relaxation rate, proving that the magnon chemical potential in YIG is slow to adapt instantaneously to the oscillating spin accumulation in Pt. These findings resolve a central question regarding the microscopic components of the spin transfer in SMR at ultrafast timescales.

## Results

To demonstrate the frequency-dependent SMR, we prepared the model SMR system, the YIG/Pt heterostructure. Our SMR samples are bilayers of 630-nm-thick $Y_3Fe_5O_{12}$ (YIG) capped with Pt with thickness $d_{\mathrm{HM}} = 0, 3, 5\ \mathrm{nm}$, where $d_{\mathrm{HM}} = 0\ \mathrm{nm}$ is a control sample.

### DC & THz measurements

We begin by measuring the SMR in the DC regime in the standard experimental setup (Fig. 1b), which is equipped with a vector magnet (1 T) [7,9,11]. As shown in Fig. 2a the dependence of $R(\alpha)/\overline{R}$ on angle $\alpha$ exhibits the characteristic $\cos 2\alpha$ dependence of the SMR with a contrast $(\Delta R/\overline{R})_{DC} = 4.2 \times 10^{-4}$ for the YIG|Pt(5 nm) sample, as obtained by fitting Eq. (1). Consistent with literature we observe the enhanced SMR in YIG|Pt(3 nm) (Figure S1) and only a negligible proximity effect of a magnitude $< 1.5 \times 10^{-5}$ was found in the studied YIG|Pt samples (Figure S1).

After the DC characterization we proceed with the THz measurements which are performed in an analogous geometry shown in Fig. 1c. Here, the driving electric field is provided by the electric field of incident THz pulses and the resistance is inferred from the THz pulses transmitted through the sample and measured as a signal $S(t,\alpha)$ as a function of time $t$ and $\alpha$. Here, $\alpha$ is the angle between the polarization of the THz pulse (THz electric field $\boldsymbol{E}_{\mathrm{THz}}$) and the orientation of $\boldsymbol{M}$ (Fig. 1c), which is set by external magnetic field.

By Fourier transformation of the THz signal $S(t)$, we obtain its complex-valued Fourier amplitude $S(\omega)$ and calculate the THz sheet resistance $R(\omega)$ of the sample stack by the Tinkham formula [31,32,33] according to

$$R(\omega) = \frac{Z_0}{n(\omega)+1} \frac{1}{S_{\mathrm{ref}}(\omega)/S(\omega) - 1}. \tag{2}$$

Here, $Z_0 \approx 377\Omega$ is the vacuum impedance, and $n(\omega)$ is the refractive index of the substrate (see Methods for validity checks) and $S_{\mathrm{ref}}(\omega)$ is corresponds to the spectrum of transmitted THz waveform through the reference YIG sample.

To directly compare our THz SMR with the DC measuremenst we estimate the frequency-averaged amplitude of $R(\alpha,\omega)/\overline{R}(\omega)$, we approximate $S_{\mathrm{ref}}(\omega)/S(\omega)$ in Eq. (2) by the ratio of the peak amplitudes of the time-domain waveforms $S_{\mathrm{ref}}(t)$ and $S(t)$. The resulting $\left(R(\alpha)/\overline{R}\right)_{\mathrm{THz}}$ is displayed for YIG|Pt(5 nm) in Fig. 2a. We observe a remarkable difference between the DC and THz SMR experiments in the YIG|Pt(5 nm) – the SMR response $\left(\Delta R/\overline{R}\right)_{\mathrm{THz}} = 3.1 \times 10^{-5}$ of YIG|Pt(5 nm) is one order of magnitude smaller than the DC value (Fig. 2a,b), suggesting a pronounced frequency dependence of the SMR.

To confirm that the observed discrepancy does not originate simply from differences between the DC and THz experimental setups, we also performed the AMR test measurements in both the DC and THz regimes. Conveniently, AMR is not expected to exhibit significant frequency dependence between DC and frequencies bellow 2 THz since the Drude-like roll-off appears only above the scattering rate ($\omega/2\pi > 5\ \mathrm{THz}$ for typical ferromagnetic metals) [33]. For these measurements, we prepared reference samples $Co_{50}Fe_{50}$(10 nm) and $Ni_{81}Fe_{19}$(8 nm) which were characterized using the DC setup. Figure 2c displays the measured $R(\alpha)/\overline{R}$ for the CoFe reference sample (the corresponding data for the NiFe sample are shownin Fig. S1), exhibiting the characteristic $\cos 2\alpha$ dependence of AMR. The corresponding THz measurement is shown in the same figure (blue line). As expected, the DC and THz signals for the CoFe sample are in good agreement.

**Spectral composition of SMR**

To further explore the distinc behaviour of SMR, $R(\alpha,\omega)/\bar{R}(\omega)$ needs to be studied in a frequency-resolved experiment with a sufficient signal-to-noise ratio. To this end, we use a second THz setup in dry-air atmosphere that is driven by low-noise femtosecond laser and we harmonically switch the sample magnetization between the states $\alpha = 0°$ and $90°$ by an AC electromagnet [33,34], cf. Fig. S2. Lock-in-type detection yields the differential signal $\Delta S(t) = S(t,0°) - S(t,90°)$, complemented with the average, magnetization-independent signal $\bar{S}(t) = [S(t,0°) + S(t,90°)]/2$.

Fig. 3 shows $\Delta S$ and $\bar{S}$ in the time (Fig. 3a) and frequency domain (Fig. 3b) for CoFe(10 nm) and YIG|Pt(5 nm). For the CoFe reference sample, the signals $\Delta S(t)$ and $\bar{S}(t)$ have a nearly identical shape (Fig. 3a). This behavior is consistent with a nearly frequency-independent response of the AMR, as observed previously for Ni, Co, $Ni_{81}Fe_{19}$ and $Ni_{50}Fe_{50}$ in the same frequency range 0-3 THz [33,35]. In stark contrast, the differential signal $\Delta S(t)$ from YIG|Pt(5 nm) has a much smaller amplitude than $\Delta S(t)$ from CoFe and oscillates considerably slower (Fig. 3a). Indeed, the amplitude spectrum $|\Delta S(\omega)|$ is concentrated at lower frequencies (Fig. 3b) as compared to $|\bar{S}(\omega)|$. This observation suggests that a significantly slower process is involved in the SMR of YIG|Pt(5 nm) than in the AMR of CoFe(10 nm), which is governed by the electron momentum relaxation with a rate of several 10 THz [33,35]. The very same trend is observed in YIG|Pt(3 nm) (see Fig. S3). Importantly, the signal $\Delta S(t)$ from the reference YIG sample without platinum is smaller than our detection sensitivity, thereby confirming that the origin of the signal is at the YIG/Pt interface (Fig. S3). This is consistent with no expected response from the solely YIG layer as its resonances are situated outside our frequency range (optical phonons above 3 THz [36], or optical magnons above 5 THz [37]). The absence of these resonances can be also seen in the spectrum of THz conductivity (Fig. S4).

To infer the THz magnetoresistive contrast $(\Delta R/\bar{R})(\omega)$ vs frequency $\omega/2\pi$ from the data in Fig. 3, we use the relation [33]

$$\frac{\Delta R}{\bar{R}}(\omega) = \frac{\Delta S(\omega)}{S(\omega)}\left[1 + \frac{n(\omega)+1}{Z_0/\bar{R}\,(\omega)}\right], \quad (3)$$

where $\bar{R}$ is the mean THz resistance of the samples as determined by an additional experiment (see Fig. S4). The magnitude of the resulting THz and DC magnetoresistive contrast $(\Delta R/\bar{R})(\omega)$ is displayed in Fig. 4 for all samples. As expected from Fig. 3, the AMR is virtually frequency-independent, and its magnitude agrees well with the DC value. In striking contrast, in both YIG|Pt samples, we observe a significant drop of the SMR contrast by more than 75% between 0 THz and 0.2 THz, followed by a further decrease down to the noise floor at 1.5 THz.

## Modelling and Discussions

The low-pass-filter type behavior of the SMR contrast $\Delta R/\bar{R}(\omega)$ (Fig. 4) must be related to a physical process that is slow compared to the period $2\pi/\omega \sim 1$ ps of our harmonic driving. To obtain more insight, we follow the theory of Refs. [23,24], were the SMR is linked to the reduction of the transient spin accumulation $\boldsymbol{\mu}_s$ in Pt at the Pt/YIG interface due to a spin current $\boldsymbol{j}_s$ from Pt to YIG (Fig. 1a) [5,7,13]. The outflow can be quantified by effective spin impedances $Z_\parallel = \mu_{s\parallel}/j_{s\parallel}$ and $Z_\perp = \mu_{s\perp}/j_{s\perp}$. Here, $\mu_{s\parallel}$ and $\mu_{s\perp}$ is the component of $\boldsymbol{\mu}_s$ parallel (∥) and perpendicular (⊥) to $\boldsymbol{M}$, respectively, and $j_{s\parallel}$ and $j_{s\perp}$ are defined likewise. We note that the perpendicular components $j_{s\perp}$, $\mu_{s\perp}$, and $Z_\perp$ are complex, encoding the two spatial directions perpendicular to $\boldsymbol{M}$ [23,24]. The SMR contrast [Eq. (1)] is proportional to the difference of the outflows for the cases $\boldsymbol{M} \perp \boldsymbol{\mu}_s$ (i.e., $\alpha = 0°$) and $\boldsymbol{M} \parallel \boldsymbol{\mu}_s$ ($\alpha = 90°$), i.e.,

$$\frac{\Delta R}{\bar{R}}(\omega) = \xi \mathrm{Re}\left[\frac{1}{Z_\perp(\omega)} - \frac{1}{Z_\parallel(\omega)}\right], \quad (4)$$

where $\xi$ is a frequency-independent coefficient (see comments in Supp. Note 1 in Supporting information or [23,24]). Because the observed resistance is higher for $\alpha = 0°$ than for $\alpha = 90°$ (Fig. 2c) and, thus, $(\Delta R/\overline{R})(\omega) > 0$, Eq. (4) implies that the transverse spin channel is dominant, i.e., $1/Z_\perp > 1/Z_\parallel$.

**Transverse coherent torque**

Therefore, we first consider solely the transverse case (⊥) where the spin outflow is due to coherent spin-torque interaction of $\boldsymbol{\mu}_s$ and $\boldsymbol{M}$. Here, the total outflow is limited by two subsequent processes (⊥1) and (⊥2). For (⊥1), the transmission of the angular momentum from platinum through the interface into YIG is described by the spin-mixing conductance, $g_{\uparrow\downarrow} \propto 1/Z_{\perp,\mathrm{FI/HM}}$ similar to spin-pumping and spin-transfer-torque experiments [38]. The transfer is instantaneous since it is a local interaction where no magnetization dynamics is involved. For (⊥2), the actual temporal dynamics of $\boldsymbol{M}$ carries the angular momentum away from the interface into the FI bulk. This process is described by the Landau-Lifshitz-Gilbert equation, and its efficiency scales as $1/Z_{\perp,\mathrm{FI}} \propto \omega^{-1/2}$ for $\omega \gg \omega_0$, where $\omega_0/2\pi \approx 1.7\,\mathrm{GHz}$ is the ferromagnetic resonance frequency of YIG [39]. This behavior is typical for oscillators driven at frequencies above resonance because the oscillator response lags behind the driving force with little transfer of angular momentum. As the processes (⊥1) and (⊥2) occur sequentially the total impedance $Z_\perp(\omega)$ equals the sum $Z_{\perp,\mathrm{FI/HM}} + Z_{\perp,\mathrm{FI}}(\omega)$ with $Z_{\perp,\mathrm{FI}}(\omega) \propto \omega^{+1/2}$ for $\omega \gg \omega_0$. We note that we do not consider the direct influence of the magnetic field component of the THz pulse, as this process is efficient only near the ferromagnetic resonance frequency, which lies far bellow our spectral range [23]. Consistently, we see no resonance-like feature in our THz transmission (see Fig. S4).

To evaluate the impact of $Z_{\perp,\mathrm{FI}}(\omega)$ on the SMR, we model the SMR with Eq. (4) and neglect the longitudinal (∥) spin channel, i.e., $Z_\parallel(\omega) \to \infty$. We take the explicit formulation of $Z_\perp$ from [23] with reported values of the saturation magnetization $M_s \approx 140\,\mathrm{kA/m}$ for YIG [40], the spin Hall angle $\theta_{HM} \approx 0.1$ for platinum [7,40] and the measured THz sheet resistance $R_{\mathrm{HM}} = 83\,\Omega$ of the platinum layer using Eq. (2). The spin stiffness $D_{\mathrm{ex}}$ and the Gilbert damping $\alpha_{\mathrm{G}}$ of YIG, the spin-accumulation relaxation length $\lambda_{\mathrm{HM}}$ of platinum, and $g_{\uparrow\downarrow}$ of the FI/HM interface are kept as free parameters whose values lie in the ranges reported in the literature (see Tab. S1). The measured DC value of the SMR contrast is taken as boundary condition. As seen in Fig. 4(b) (grey curves and the shaded area between them), the modeled SMR contrast and, thus, the transverse component $Z_\perp(\omega)$ do not capture the strong frequency dependence of the measured $(\Delta R/\overline{R})(\omega)$. Therefore, we focus on the longitudinal SMR contributions next.

**Longitudinal incoherent torque**

The longitudinal component of SMR, $Z_\parallel(\omega) = Z_{\parallel,\mathrm{FI/HM}} + Z_{\parallel,\mathrm{FI}}(\omega)$, is also formed by two sequential processes (∥1) and (∥2) creating incoherent (thermal) spin fluctuations. In (∥1), spin is transmitted through the interface. This process is instantaneous (compared to our time scales) and, thus, implies a frequency-independent $1/Z_{\parallel,\mathrm{FI/HM}}$. In (∥2), the angular momentum is carried away from the interface by magnons. The efficiency of this process scales with the gradient of the magnon chemical potential $\mu_{\mathrm{m}}$ in FI and, thus, with the inverse diffusion length $1/\lambda_{\mathrm{m}}$ of the magnon population, i.e., $1/Z_{\parallel,\mathrm{FI}}(\omega) \propto 1/\lambda_{\mathrm{m}}(\omega)$. By considering the equation of motion for $\mu_{\mathrm{m}}$ (see Supp. Note 2), one can show that

$$\lambda_{\mathrm{m}}^2(\omega) \propto \frac{\tau_{\mathrm{m}}^{\mathrm{mom}}\tau_{\mathrm{m}}^{\mathrm{pop}}}{(1 - \mathrm{i}\omega\tau_{\mathrm{m}}^{\mathrm{mom}})(1 - \mathrm{i}\omega\tau_{\mathrm{m}}^{\mathrm{pop}})}. \quad (5)$$

Here, $\tau_{\mathrm{m}}^{\mathrm{mom}}$ and $\tau_{\mathrm{m}}^{\mathrm{pop}}$ is, respectively, the relaxation time of magnon momentum and magnon population. By considering that, typically, $\tau_{\mathrm{m}}^{\mathrm{mom}} \sim 0.1$ ps [25] and $\tau_{\mathrm{m}}^{\mathrm{pop}} \sim 100$ ps [24] to 100 ns [41], Eq. (5) yields the approximate scaling $Z_{\parallel,\mathrm{FI}}(\omega) \propto \lambda_{\mathrm{m}}(\omega) \propto \omega^{-1/2}$ for $10\,\mathrm{GHz} < \omega < 10\,\mathrm{THz}$ (see Supp. Note 2 in Supporting information). Therefore, Eq. (5) can explain the observed frequency dependence of the SMR (Fig. 4b).

To include this process in our model, we fit the whole Eq. (4) to our data (Fig. 4b) with the same conditions as above. We obtain excellent agreement of measured data and fit (Fig. 4b) in the covered frequency window 0-1.5 THz, and all fit parameters agree well with the ranges reported in literature or are close to it: $D_{\mathrm{ex}} = 4 \times 10^{-6}\ \mathrm{m}^2\ \mathrm{s}^{-1}$ [42], $\alpha_{\mathrm{G}} = 1.6 \times 10^{-3}$ [43], $\lambda_{\mathrm{HM}} = 1.1\ \mathrm{nm}$ [44], $g_{\uparrow\downarrow} = (3.4 + 0.2\mathrm{i}) \times 10^{14}\ \Omega^{-1}\ \mathrm{m}^{-2}$ [7] (see Tab. S1). This agreement indicates that our model captures the relevant contributions to the THz SMR in YIG|Pt.

**Discussions**

The model implies that the observed decrease of the SMR contrast with increasing frequency is the result of a competition [Eq. (4)] of the transverse, nearly frequency-independent contribution $1/Z_{\perp}(\omega)$ and the longitudinal, frequency-dependent component $1/Z_{\parallel}(\omega)$ that increases with $\omega$ (see Fig. 4c). At about 1.5 THz, the two contributions approximately cancel each other. As both $Z_{\perp}$ and $Z_{\parallel}$ depend on different material parameters, the SMR contrast may even become negative at frequencies above 1.5 THz. A change in material parameters, temperature or magnetic field may shift the frequency region where both components cancel.

The observed low-pass behavior in the THz SMR cannot be explained by a hypothetical proximity-effect-induced AMR, because the vanishing of the THz SMR above 1.5 THz implies that the decay would have to originate from extrinsic THz AMR following the Drude-like THz conductivity of Pt [33]. However, the extracted THz conductivity of Pt is practically constant in our samples (Fig. S5), thereby excluding a proximity-effect origin of these results.

The $\lambda_{\mathrm{m}}$ vs $\omega$ resulting from the model (solid curve in Fig. 4d) indicates that incoherent magnons in YIG have diffusion lengths in the nanometer range above 0.2 THz and that the DC limit of $\lambda_{\mathrm{m}}$ ($\approx 2\ \mathrm{nm}$) is significantly smaller than the hundreds of nanometers or even micrometers inferred from DC experiments [25,27,28,29,30]. We attribute this underestimation to the relaxation-time approximation of magnon scattering processes used in the model [24]. To confirm that our conclusion is still valid for larger DC relaxation lengths, we fix $\tau_{\mathrm{m}}^{\mathrm{pop}}$ to milliseconds to reproduce the most extreme case of $\lambda_m(\omega = 0) = 9\ \mu\mathrm{m}$ [25] (see Supp. Note 2 in Supporting information), and plotted the fits as dashed curves in Fig. 4b-d. This procedure yields similar results in the THz frequency range for fitting parameters still well within the literature values or close to them (Tab. S1). In fact, this behavior is expected from Eq. (5) as $\lambda_{\mathrm{m}}$ is independent of $\tau_{\mathrm{m}}^{\mathrm{pop}}$ for $\omega \gg 1/\tau_{\mathrm{m}}^{\mathrm{pop}}$. Similarly, because of this, our model and corresponding conclusions for THz frequencies are robust against small changes of DC SMR value [see Fig. S5(b)], for example due to sample inhomogeneity as reported in literature [45]

The observed low-pass spectral filtering of THz SMR provides a versatile fingerprint expected across a wide range of magnetic systems, with the cut-off frequency being highly sensitive to material parameters. This makes THz SMR an ideal diagnostic tool for optimizing spintronic interfaces. For instance, while high conductivity in metallic ferromagnets may suppress the incoherent magnon contribution due to spin transport by conduction electrons, materials with low saturation magnetization $M_s$ and large spin-mixing conductance $g_{\uparrow\downarrow}$, like ferrimagnetic Heusler compounds, are predicted to exhibit significantly enhanced spectral filtering due to efficient incoherent magnon transport [23].

Most crucially, our approach is pivotal for the future antiferromagnetic and altermagentic spintronic research.Because many antiferromagnetic and altermagnetic materials are insulators, SMR is particularly useful for probing spin dynamics at their interfaces.Given that the intrinsic magnon frequencies in these materials are inherently in the THz bandwidth, THz SMR becomes the essential, non-contact tool for directly accessing their unique ultrafast spin-magnon coupling, opening the door to high-speed device concepts.

In conclusion, we report the first experimental observation of the SMR effect bridging the gap between static transport and the ultrafast regime (0–1.5 THz) in the prototypical YIG|Pt system. We observe a dramatic spectral decay, where the SMR contrast drops by more than an order of magnitude at terahertz frequencies. By combining these measurements with a rigorous dynamic model, we resolve the microscopic origin of this behavior: it arises from a competition between transverse (coherent) and longitudinal (incoherent) spin-torque contributions. Specifically, we identify that the efficient flow of incoherent magnons, driven at higher frequencies than their population relaxation rates, leads to a partial cancellation of the total contrast in spin currents through the interface. These insights not only provide an insightful physical picture of interfacial spin transport but also establish THz SMR as a powerful, non-contact probe of ultrafast spin-magnon coupling, which can be in particularly useful to explore high-frequency transport in insulating antiferromagnets and the emerging class of altermagnets.

## Methods

### Samples and growth

The SMR samples were grown by liquid-phase epitaxy of $Y_3Fe_5O_{12}$ (YIG) on (111)-oriented single-crystalline gadolinium garnet substrates (GGG), followed by a deposition of platinum by sputtering after removing organic residues by a solution of sulfuric and hydrogen peroxide. The resulting stack is GGG|YIG($d_{\mathrm{FI}}$)|Pt($d_{\mathrm{HM}}$) with $d_{\mathrm{FI}} = 630\ \mathrm{nm}$ and $d_{\mathrm{HM}} = 0, 3, 5\ \mathrm{nm}$, where $d_{\mathrm{HM}} = 0\ \mathrm{nm}$ is a control sample. The reference AMR samples are single ferromagnetic layers of single-crystalline $Co_{50}Fe_{50}$(10 nm) and polycrystalline $Ni_{81}Fe_{19}$(8 nm), prepared by molecular-beam epitaxy on a MgO(100) substrate and by sputtering on a high-resistivity silicon substrate, respectively.

### THz experiments

Spectrally unresolved experiment: Pulses of THz radiation are generated by optical rectification of femtosecond laser pulses from a regenerative amplifier (wavelength 1030 nm, duration 170 fs, energy 50 μJ, repetition rate 10 kHz) in a GaP(110) crystal (thickness 2 mm). After transmission through the sample, the THz electric field is characterized by electro-optic sampling (EOS) in GaP(110) (2 mm). The external magnetic field is provided by a rotatable Halbach array of amplitude ≈350 mT.

Spectrally resolved experiment: A low-noise oscillator laser (wavelength 800 nm, duration 10 fs, energy 2 nJ, repetition rate 80 MHz) is used to pump a passive photoconductive switch (Terablast, Protemics GmbH) and the EOS is done using a ZnTe(110) (1 mm) crystal. The harmonic magnetic field is delivered an AC electromagnet of peak amplitude 20 mT and frequency 10 kHz [33,34].

### Validity of Tinkham formula

The transmission coefficient $t(\omega)$ for a thin film deposited in the substrate yields

$$t(\omega) = \frac{E(\omega)}{E_{ref}(\omega)} = \frac{2n(n_s+1)\exp\left(\frac{-i\omega(n-1)d}{c}\right)\exp\left(\frac{-i\omega(d_s-d_r)(n_s-1)}{c}\right)}{(n+1)(n_s+n)-(n-1)(n-n_s)\exp\left(\frac{-2i\omega nd}{c}\right)},$$

where $E(\omega)$ and $E_{ref}(\omega)$, are respectively the spectrum of THz wave transmitted through the thin film and the substrate, $n$ and $n_s$ correspond to refractive index of the thin film (Pt layer) and substrate (GGG, $n_s = 4.1$) [46], $d$ is the thickness of the thin film and $d_s$ and $d_r$ thicknesses of the substrate below the Pt layer and the reference substrate, and $c$ is the speed of light in vacuum. In the first order of expansion in $nd$ (the thin film approximation), it follows

$$t(\omega) = \frac{n_s+1}{n_s+1+a+Z_0\sigma d}\exp(-i\omega \Delta d\,(n_s-1)/c),$$

where $Z_0$ and $\sigma$ are the vacuum impedance and THz conductivity of the thin film, and $a$ equals to $\omega n_s d/c$. In order to use the Tinkham formula, two conditions need to hold: (i) $n \gg n_s$ and (ii) $a \ll 1$. Assuming no interband transitions in the thin film (yielding the lower estimate for $n$), the refractive index can be estimated as $n = \sqrt{1 - i\,\omega/\sigma\,\epsilon_0}$ . Using the conductivity of platinum in YIG|Pt(5 nm), $\sigma = 2.4 \times 10^6 Sm^{-1}$, the refractive index varies from $n = 288 - 288i$ at 0.2 THz to $n = 94 - 94i$ at 2 THz, which is consistent with literature values [47], safely larger than $n$ of YIG in our spectral range and verifies condition (i). Condition (ii) is verified by calculating $a$ for the maximal film thickness $d$ = 5 nm, yielding $a$ < 0.0013 for studied spectral range, which fulfils the condition.

### Data availability

The datasets generated and/or analyzed during the current study are available in the Zenodo data repository at the address https://doi.org/10.5281/zenodo.20345696.

## Acknowledgments

We thank Martin Wolf and the Fritz Haber Institute in Berlin, Germany for their support in the initial phase of our experiments. This work was supported by TERAFIT project No. CZ.02.01.01/00/22_008/0004594 funded by Ministry of Education Youth and Sports of the Czech Republic (MEYS CR), programme Johannes Amos Comenius (OP JAK), call Excellent Research. The authors acknowledge funding by the Czech Science Foundation (grant no. 21–28876J), by the Grant Agency of the Charles University (grants no. 166123, no. 120324 and SVV–2025–260836), by CzechNanoLab Research Infrastructure supported by MEYS CR (LM2023051), by MEYS CR project LNSM-LNSpin, by the German Research Foundation through the collaborative research center CRC/TRR 227 “Ultrafast spin dynamics” (project ID 328545488, projects A05, B02 and B03), the research unit RU 5844 “Chiral phonons in spintronics” (project KA 3305/7-1) and the European Research Council through the ERC AdG ORBITERA (grant nr. 101142285 ORBITERA), the Deutsche Forschungsgemeinschaft (DFG, German Research Foundation) through TRR 173/3 Spin+X (project ID 268565370, projects B02, B13, and A01) and SFB 1432 (project ID 425217212), and the European Union through the Horizon 2020 and HorizonEurope projects NIMFEIA (grant no. 101070290) EIC PathfinderOPEN grant OBELIX (grant no. 101129641) and the Horizon Europe programme under Grant Agreement No. 101211111 (Tera-MaRs). Y. Wu acknowledges the National Key Research and Development Program of China (Grant No. 2024FYA1408500). M. L. is cofunded by the European Union (Physics for Future – Grant Agreement No. 101081515).

*Supporting information for*

# Observation of Ultrafast Coherent and Incoherent Spin Torques via Terahertz Spin-Hall Magnetoresistance

Peter Kubaščík[1], Richard Schlitz[2], Oliver Gueckstock[3], Oliver Franke[3], Miina Leiviskä[4], Martin Borchert[5], Gerhard Jakob[6], Kamil Olejník[4], Andrej Farkaš[1,4], Zdenek Kašpar[1,4], Jiří Jechumtál[1], Martin Bušina[1], Eva Schmoranzerová[1], Petr Němec[1], Yicheng Z. Wu[7,8], Georg Woltersdorf[9], Mathias Kläui[6], H. Reichlova[4], Piet W. Brouwer[3], Sebastian T. B. Goennenwein[2], Tobias Kampfrath[3], Lukáš Nádvorník[1,*]

1. Faculty of Mathematics and Physics, Charles University, 121 16 Prague, Czech Republic
2. Department of Physics, University of Konstanz, Konstanz, Germany
3. Department of Physics, Freie Universität Berlin, 14195 Berlin, Germany
4. Insitute of Physics, Czech Academy of Sciences, Prague, Czech Republic
5. Max Born Institute for Nonlinear Optics and Short Pulse Spectroscopy, 12489 Berlin, Germany
6. Institut für Physik, Johannes Gutenberg-Universität Mainz, 55128 Mainz, Germany
7. Department of Physics, State Key Laboratory of Surface Physics, Fudan University, Shanghai 200433, China
8. Shanghai Research Center for Quantum Sciences, Shanghai 201315, China
9. Institut für Physik, Martin-Luther-Universität, Halle, Germany

[*] E-mail: lukas.nadvornik@matfyz.cuni.cz

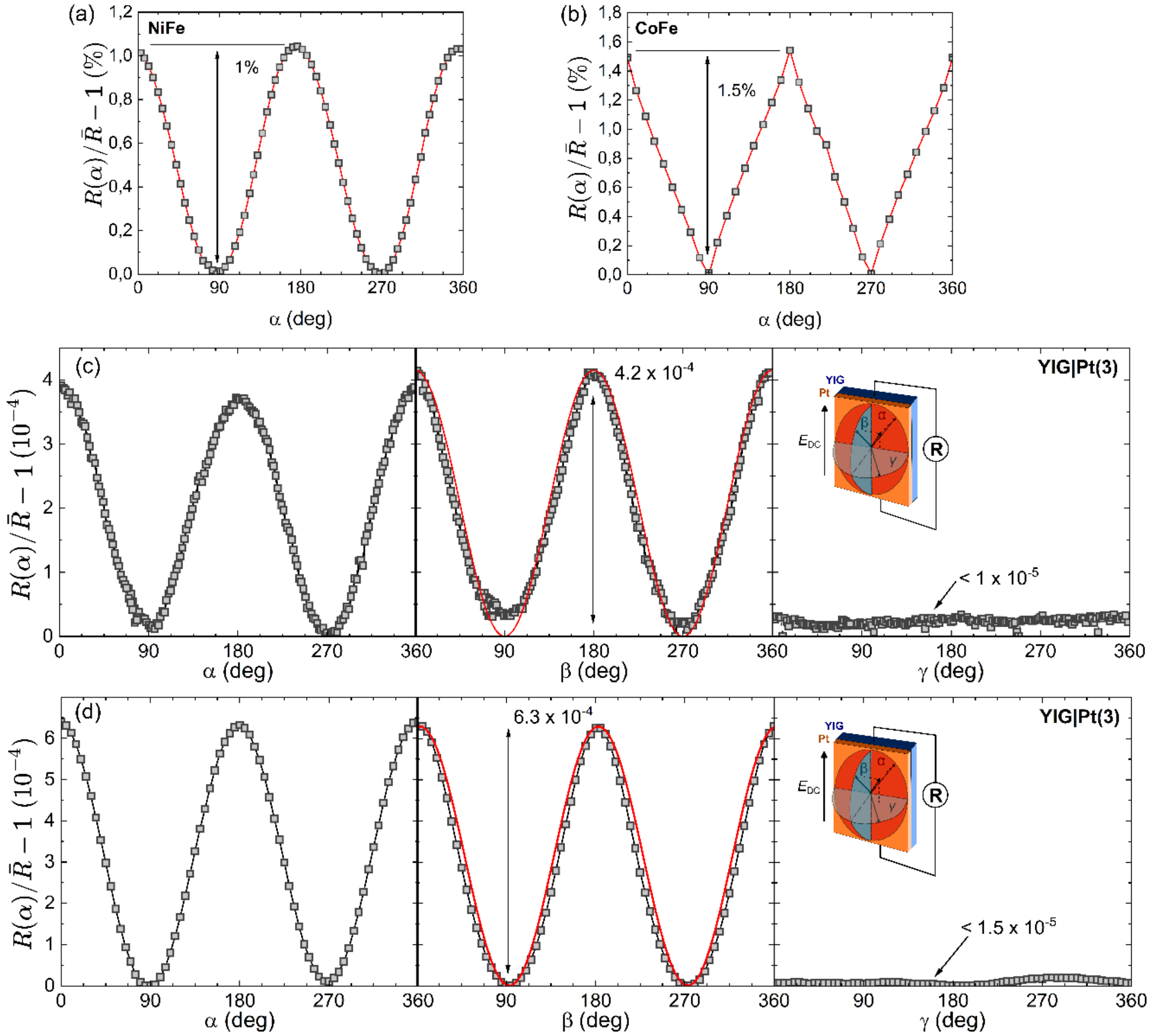

**Figure S1**. DC SMR in presented samples. Scatters correspond to the relative change of resistance $R(\alpha)/\bar{R}-1$ of (a) NiFe (b) CoFe (c)YIG|Pt(3) and (d) YIG|Pt(5). Amplitudes of DC AMR(SMR) contrasts are depicted by double-side arrows. Data in (b) were measured with external field of 30 mT to consider a not fully saturated magnetization under these conditions. All other samples were well saturated with 30 mT (see Fig. S2 for YIG and Ref[1]. for NiFe). In case of YIG|Pt(3) and YIG|Pt(5) the relative change of resistance was measured also in out-of-plane directions in order to exclude proximity effect in Pt layer (for more details see Ref [2]).

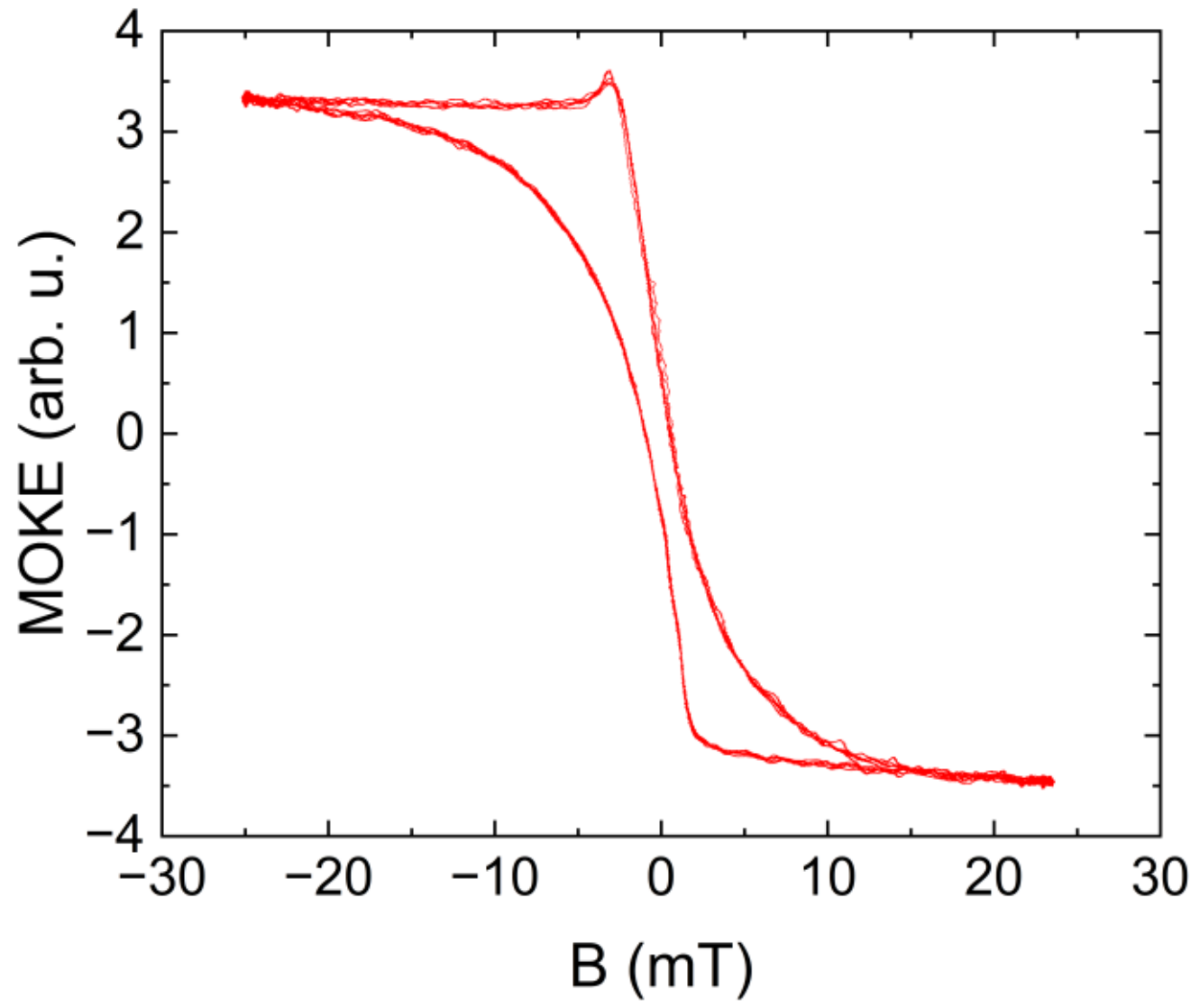

**Figure S2.** Hysteresis loop measured by magnetooptical Kerr effect in YIG/Pt(5). The loop indicates the saturation field to be around 20 mT.

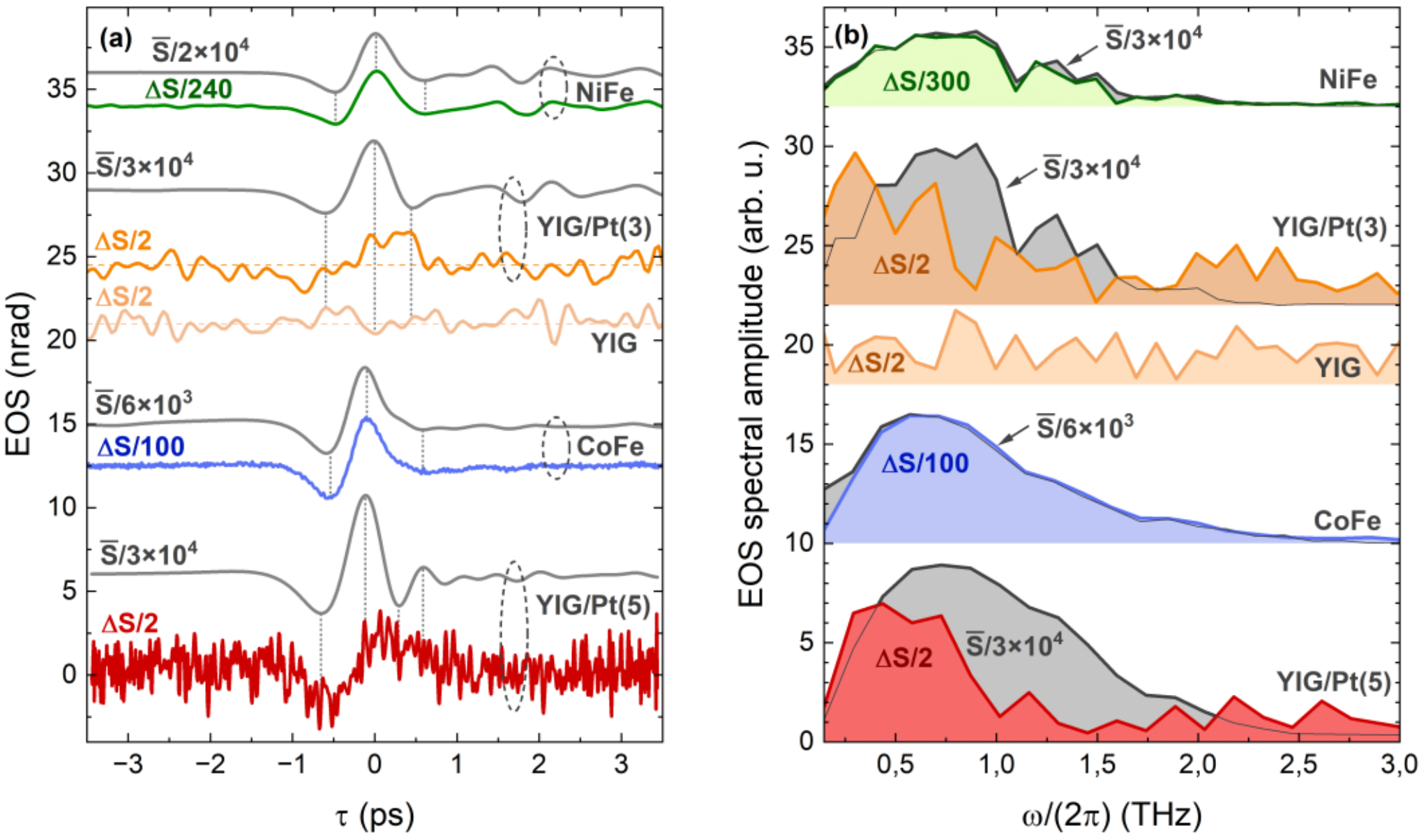

**Figure S3.** All measured terahertz signals and corresponding spectra. (a) Measured THz waveforms of the magnetization-independent average transmission $\overline{S} = (S_{\parallel} + S_{\perp})/2$ (grey curves, rescaled by indicated factor) and the magnetization-dependent differential signals $\Delta S = S_{\parallel} - S_{\perp}$ for YIG/Pt(3 nm) (orange curve), YIG/Pt(5 nm) (red) and references NiFe (green) and CoFe samples (blue curve). Dotted lines are guide for the eye. (b) Spectra corresponding to (a), rescaled indicated factors. Curves are shifted vertically for clarity.

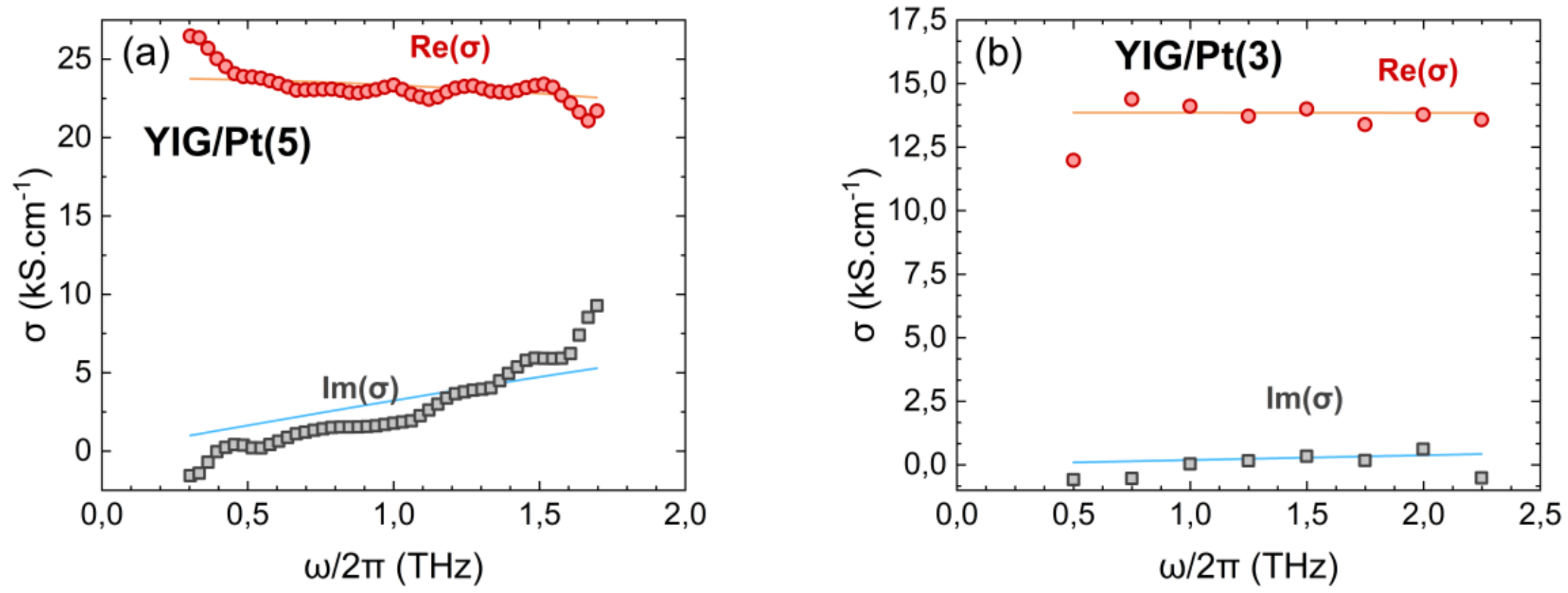

**Figure S4.** THz conductivity of YIG/Pt(5 nm) and YIG/Pt(3 nm). Red dots (black squares) shows real (imaginary) part of THz conductivity and orange (blue) fit by Drude model of conductivity $\sigma_0/(1 - i\omega\tau_k)$, where $\sigma_0$ is the DC conductivity and $\tau_k$ is the electron momentum relaxation time. The conductivities of NiFe and CoFe are frequency-independent in our spectral range and reach $\sigma \approx 2.9MSm^{-1}$ and $7.1MSm^{-1}$, respectively [1].

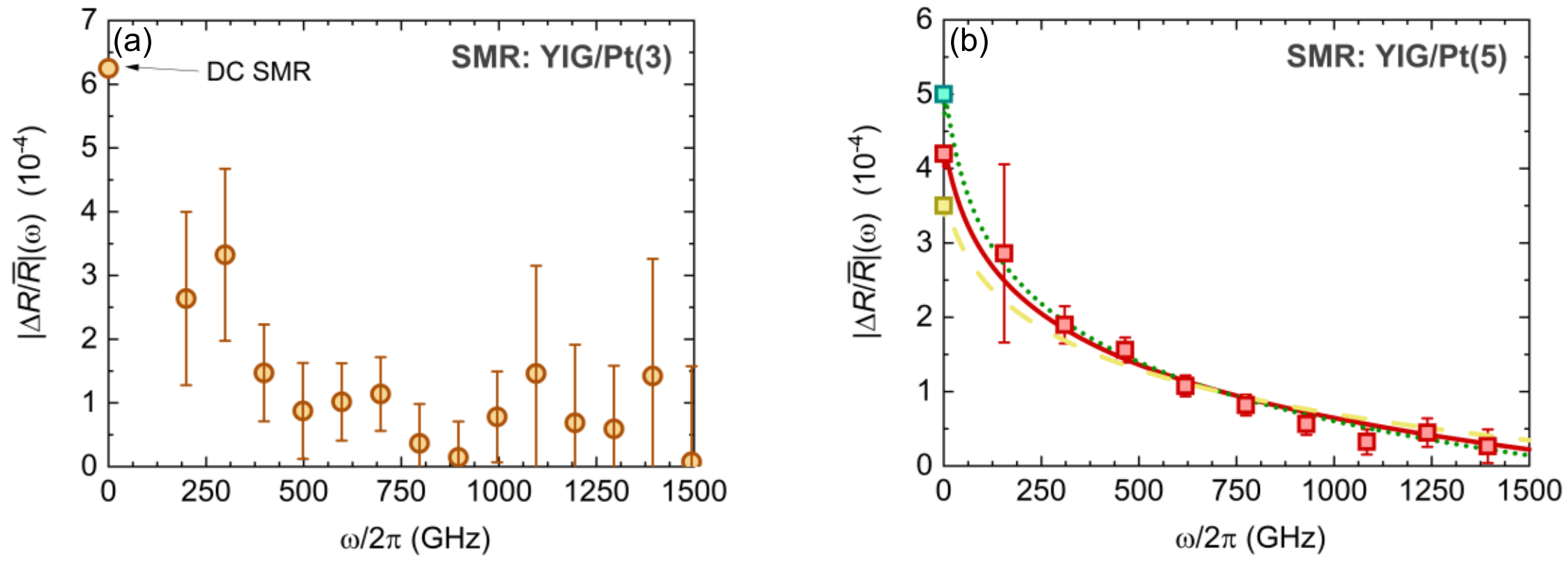

**Figure S5**. (a) The DC and THz SMR measured in YIG|Pt(3) sample. (b) Comparisson of the SMR for different DC SMR ratios. The green and yellow points corresponds to upper and lower bound of DC SMR ratio and yellow dashed and green dot curve shows the modeled SMR profile. As we can see, the DC SMR value impacts the model only at low frequencies, while at THz frequencies the model converges to the same results. The values of the free material parameters used in the model of the THz spectrum of the SMR ratio are, for the upper bound of the DC SMR ratio, $D_{\mathrm{ex}} = 4.4 \times 10^{-6}\ \mathrm{m}^2\ \mathrm{s}^{-1}$, $\alpha_{\mathrm{G}} = 1.6 \times 10^{-3}$, $\lambda_{\mathrm{HM}} = 1.1\ \mathrm{nm}$, $g_{\uparrow\downarrow} = (4.2 + 0.3\mathrm{i}) \times 10^{14}\ \Omega^{-1}\ \mathrm{m}^{-2}$ , and, for the lower bound of the DC SMR ratio, $D_{\mathrm{ex}} = 4 \times 10^{-6}\ \mathrm{m}^2\ \mathrm{s}^{-1}$, $\alpha_{\mathrm{G}} = 1.9 \times 10^{-3}$, $\lambda_{\mathrm{HM}} = 1\ \mathrm{nm}$, $g_{\uparrow\downarrow} = (3.4 + 0.2\mathrm{i}) \times 10^{14}\ \Omega^{-1}\ \mathrm{m}^{-2}$.

**Full model**

| **Fixed parameter** | **Value** | | |
|---|---|---|---|
| saturation magnetization of YIG, $M_s$ | 140 kA/m [3] | | |
| spin Hall angle of Pt, $\theta_N$ | 0.1 [2] | | |
| resistance of Pt, $R_N$ | 83 Ω (measured) | | |
| **Free parameter** | **Range of literature values** | **Fitted values by Eq. (4)** | **Fitted values by Eq. (4) with fixed DC limit of $\lambda_m$** |
| spin stiffness of YIG, $D_{ex}$ | $4\times10^{-6}$ – $1.1\times10^{-5}$ $m^2\ s^{-1}$ [3], [4], [5], [6] | $4\times10^{-6}$ $m^2\ s^{-1}$ | $4\times10^{-6}$ $m^2\ s^{-1}$ |
| Gilbert damping of YIG, $\alpha_G$ | $7\times10^{-5}$ – $1\times10^{-3}$[7], [8], [9], [10], [11], [12] | $1.6\times10^{-3}$ | fixed ($2.3\times10^{-10}$) |
| spin relaxation length in Pt, $\lambda_N$ | 1 – 4 nm [2], [3], [13] | 1.1 nm | 1 nm |
| spin-mixing conductance of YIG\|Pt, $g_{\uparrow\downarrow}$ | $5\times10^{13}$ – $9\times10^{14}$ $\Omega^{-1}m^{-2}$ [2], [3], [7], [14], [15], [16], [17] | $(3.4+0.2i)\times10^{14}$ $\Omega^{-1}m^{-2}$ | $(3.2+0.2i)\times10^{14}$ $\Omega^{-1}m^{-2}$ |
| relaxation time, $\tau_m^{mom}$ | | 13 fs | 10 fs |

**Model with only coherent torque contribution**

| **Fixed parameter** | **Value** | |
|---|---|---|
| saturation magnetization of YIG, $M_s$ | 140 kA/m [3] | |
| spin Hall angle of Pt, $\theta_N$ | 0.1 [2] | |
| resistance of Pt, $R_N$ | 83 Ω (measured) | |
| Gilbert damping of YIG, $\alpha$ | $1.6\times10^{-3}$ (from fit by Full model) | |
| **Free parameter** | **Best fit (lower grey curve in Fig. 4)** | **Worst fit (upper grey curve)** |
| spin stiffness of YIG, $D_{ex}$ | $4\times10^{-6}$ $m^2\ s^{-1}$ | $1.1\times10^{-5}$ $m^2\ s^{-1}$ |
| spin relaxation length in Pt, $\lambda_N$ | 1 nm | 4 nm |
| spin-mixing conductance of YIG\|Pt, $g_{\uparrow\downarrow}$ | $9\times10^{14}$ $\Omega^{-1}m^{-2}$ | $3.7\times10^{13}$ $\Omega^{-1}m^{-2}$ |

**Table S1:** Tables of used and fitted material parameters for SMR modelling. Full model: The fixed parameters are constants used in the fits, obtained from indicated source. Fitted values are results of fits by the full Eq. (4) with calculated $\lambda_m(\omega)$ (see Note 2) either without constraints on $\lambda_m$, or with the DC value $\lambda_m(0) = 9\mu m$ taken from literature[18]. $\lambda_m(0)$ is mainly determined by $\tau_m^{pop} = \hbar/\alpha_G k_B T \approx$1 ms. We therefore indirectly set $\lambda_m(0)$ by fixing Gilbert damping $\alpha_G$ and find that the SMR contrast is relatively insensitive to $\lambda_m(0)$, $\tau_m^{pop}$, or $\alpha_G$ within our theory. Model with only coherent torque contribution ($Z_\perp$): The range of literature values was used as limits for fitting parameters with the DC value of SMR as a boundary condition, providing a range of results (grey area in Fig. 3) between the best and worst fit (lower and upper grey curves in Fig. 3).

## Supporting Notes

### Supporting Note 1: THz SMR model and modeling of THz response

In Fig. 4 of the main text, we fit the measured frequency-dependent spin-Hall magnetoresistance using the theory of Ref. [19]. We here describe how the fitted curves were obtained from the theoretical model of Ref. [19].

Reference [19] employs an effective magneto-electric circuit theory similar to that of Ref. [4] for the coupled spin and thermal response to linear order in the electric field $E(\omega)$ applied in the normal metal NM. In the circuit description, the response of each individual part of the system — the normal-metal layer NM, the magnetic layer FM, and their interface — is described with the help of an effective impedance. In a second step, the response of the total system is found using Kirchhoff-like laws to combine these impedances. The impedances relate spin and heat currents to their "driving forces", which are spin accumulation and temperature for NM and magnon chemical potential and temperature for FM[20]. Magnonic relaxation processes are included in the effective impedance of the FM layer. The circuit theory of Ref. [19] is the finite-frequency generalization of previous theories of the SMR effect in the low-frequency limit ([21], [22], [23], [24]).

In the magnet FM, the spin current $j_s$ is carried by magnons. Its components $j_{s,\perp}$ and $j_{s,\parallel}$ perpendicular and parallel to the magnetization describe qualitatively different types of magnon response: The transverse component is related to the coherent magnetization dynamics, whereas the longitudinal component describes spin transport by incoherent, thermal magnons. In Ref.[19] , the former is described using the Landau-Lifshitz-Gilbert equation, whereas the latter is described by a diffusion-type approach. The relevant parameters describing the FM layer in the theory of Ref. [19] are its thickness $d_M$, saturation magnetization $M_s$, ferrimagnetic resonance frequency $\omega_0$, spin stiffness $D_{ex}$ and Gilbert damping coefficient $\alpha$. Further, for spin and heat transport by thermal magnons, Ref. [19] invokes relaxation processes phenomenologically described by the magnon momentum relaxation time $\tau_m^{mom}$ which is dominated by elastic magnon-impurity scattering, as well as the relaxation times $\tau_{m,ex}$, $\tau_{mp,ex}$, and $\tau_m^{pop}$ for exchange-based magnon-magnon scattering, exchange-based magnon-phonon scattering and spin-nonconserving (relativistic) magnon-phonon and magnon-magnon scattering, respectively.

In Ref. [19] the NM|FM interface is described using the spin mixing conductance $g_{\uparrow\downarrow}$, which controls the coherent transverse spin current, as well as the incoherent, longitudinal spin current and the heat current through the interface ([25], [26], [27]). Transport of charge and spin in NM is coupled via the spin-Hall effect and the inverse spin-Hall effect ([28], [29]), which are described phenomenologically using the spin-Hall angle $\theta_{NM}$. Spin relaxation of conduction electrons in NM is described phenomenologically with the help of the spin relaxation length $\lambda_{NM}$. Other relevant parameters for electron transport in NM are the electrical conductivity $\sigma_{NM}$ and the thickness $d_{NM}$ of the NM layer. The theory of Ref. [19] assumes that $d_{NM} \gg \lambda_{NM}$, a condition that is satisfied in our experimental setup. Heat currents in NM are described by a diffusion equation, which includes the electronic heat capacity $C_e$, the electronic thermal conductivity $\kappa_e$, and the electron-phonon relaxation length $l_{ep}$ as its fundamental parameters. The parameter $\xi$ , which is used for converting parallel and perpendicular effective spin impedances to change in resistance, is equal to $\xi = \theta^2\hbar\sigma_N^2\bar{\rho}\, Z_N^2/4e^2 d_N$ , where $\bar{\rho}$ is average resistivity of the layer, $Z_N^2$ is effective spin impedance of HM layer and $e$ is elementary charge.

For a quantitative description of the measured SMR data, we obtain $\sigma_{NM}$ from the measured resistance of Pt(5), $R_{NM} = 83\Omega$, and we use the actual values for the thicknesses $d_{NM} = 5nm$ and $d_M = 630nm$ of the NM and FM layers. We estimate the ferrimagnetic resonance frequency $\omega_0$ by the Kittel formula, $\omega_0/2\pi = (\gamma/2\pi)\sqrt{\mu_0 H(\mu_0 H + \mu_0 M_s)} \approx 1.75GHz$, for $\mu_0 H = 20mT$ and $\mu_0 M_s = 175mT$, where we took the well-established literature value for the saturation magnetization of YIG to limit the number of free parameters ([2]). For the material parameters $\theta_{NM}$, $C_e$, and $l_{ep}$, which describe NM, we also resort to the literature values $\theta_{NM} = 0.1$, $C_e = 0.13 \times 10^6 JK^{-1}m^{-3}$, and $l_{ep} = 4.5nm$ ([2], [3], [30], [31], [32], [33]). Since the fit was found to be relatively insensitive to the relaxation times $\tau_{m,ex}$ and $\tau_{mp,ex}$, we took literature values $\tau_{m,ex} = 2.7ps$ ([34]) and $\tau_{mp,ex} = 0.5ps$ ([20]). The relaxation time for spin-nonconserving relaxation processes $\tau_m^{pop}$ is estimated as $\tau_m^{pop} \approx \hbar/\alpha_G\, k_B T$. (The same literature values and estimate were used in Ref. [19] for the numerical estimates made there.) The remaining five parameters $D_{ex}$, $\alpha_G$, $g_{\uparrow\downarrow}$, $\lambda_{NM}$, and $\tau_m^{mom}$ of the theoretical model are fitted. The fit for the relaxation time $\tau_m^{mom}$ was constrained by the condition that $80\ \mathrm{fs} < \tau_m^{mom} < \tau_{mp,el}$, where the second inequality follows from the fact that exchange-based magnon-phonon scattering also contributes to magnon momentum relaxation, so that $\tau_{mp,el}^{-1}$ is a lower bound for the magnon momentum relaxation rate[19]. The least-squares fit obtained this way very well matches the experimental data (see Fig. 4), whereas the fitted values for $D_{ex}$, $g_{\uparrow\downarrow}$, $\alpha_G$, $\lambda_{NM}$, $\tau_m^{mom}$ are within the range of values reported in the literature (see Table S1). To verify that our model remains valid for larger DC relaxation lengths than predicted by the full fit as well, we also conducted a fit with fixed $\lambda_m(0) = 9\ \mu\mathrm{m}$ , for which we refer to Note 3 and Tab. S1.

We note that the theory of Ref. [19] (as well as that of Ref. [4]) in general does not imply that $|\Delta R/\bar{R}| \rightarrow 0$ for $\omega \rightarrow \infty$. Rather, it is the difference of coherent/transverse and incoherent/longitudinal magnon transport channels that determines the high-frequency limit of the SMR effect. These depend on material properties and are expected to be different for different material combinations, but also for different temperature or a different applied magnetic field.

In the main text, we also compare our measured data with a theoretical model that includes coherent magnon transport only. This model is obtained from the theory of Refs. [4], [19] by imposing that the longitudinal spin current $j_{s\parallel}$ through the NMFM interface is zero. To obtain the curves shown in Fig. 4, we took the same value of the Gilbert damping parameter $\alpha_G$ as in the fit to the full model. Then, we varied the remaining free parameters within limit intervals with keeping the dc value as the boundary condition. The best and worst match provided by the model within the parameter space are shown by grey curves in Fig. 4 and values of their parameters are given in Table S1. Even the “best” match, which used the extremal values of the parametric space, can not explain the data. Other (non-extremal) combinations of parameters filled the space between the two extremal curves (grey-shaded area in Fig. 4).

While the DC SMR was extracted from transport measurements, the THz radiation averages over a larger area; thus, spatial variations of the DC SMR need to be addressed. To account for spatial variations of the SMR within the sample, we performed additional modeling by varying the extracted DC SMR by ±15% ($3.5 \times 10^{-4}$ to $5 \times 10^{-4}$), consistent with literature reports, without any significant change in the fit between the model and experimental data (Fig. S5).

## Supporting Note 2: Derivation of $\lambda_m(\omega)$

We express $\lambda_m(\omega)$ using the constitutive propagation equation for magnons:

$$\partial_t \mu_m - \frac{\sigma_m}{1 - i\omega\tau_m^{mom}} \nabla^2 \mu_m + \frac{\mu_m}{\tau_m^{pop}} = 0,$$

where $\sigma_m$ is the magnon conductivity, $\tau_m^{mom}$ the relaxation time of magnon momentum as obtained from Matthias rule from all contributing spin-conserving scattering processes, and $\tau_m^{pop}$ the relaxation time of the magnon population, given by the spin non-conserving scattering [20]. Rewriting the above equation into the general form of spin diffusion,

$$\nabla^2 \mu_m = \lambda_m^{-2} \mu_m,$$

yields the proportionality

$$\lambda_m^2(\omega) \propto \tau_m^{mom}\, \tau_m^{pop} / (1 - i\omega\tau_m^{mom})\left(1 - i\omega\tau_m^{pop}\right),$$

i.e., Eq. (5) of the main text. While in the DC limit, the magnon relaxation length can be approximated as $\lambda_m(0) \sim \sqrt{\tau_m^{mom}\, \hbar / \alpha_G\, k_B T}$, in the THz spectral range ($1/\tau_m^{mom} = 100\ \mathrm{THz} \gg \omega \gg \sqrt{\frac{\hbar}{\alpha_G\, k_B T}} = 4\ \mathrm{GHz}$) the magnon relaxation length is independent of $\tau_m^{pop}$ and yields $\lambda_m(\omega) \propto \sqrt{1/\omega}(1 + 1i)$, which is independent of Gilbert damping $\alpha_G$ .Since remaining model is relatively insensitive to Gilbert damping as well, we can set $\alpha_G$ outside of the range of literature values to fix the DC magnon diffusion length of 9 µm while the model remains the same in the THz frequency range. The obtained material parameters are shown in the third column in Tab. S1. Except for the Gilbert damping parameter, the resulting material parameters are again within the range of literature values.